\documentclass[%
 reprint,
 amsmath,amssymb,nofootinbib,
 aps,
]{revtex4-2}

\usepackage{graphicx}
\usepackage{bm}

\usepackage{fancyhdr}
\usepackage[colorlinks]{hyperref}
\usepackage[nameinlink,noabbrev]{cleveref}
\hypersetup{
    colorlinks=true,
    linkcolor=blue,
    filecolor=magenta,      
    urlcolor=blue,
   citecolor=blue
}
\usepackage{mathrsfs}

\usepackage{subfig}

\newsavebox{\measurebox}

\usepackage{scalerel}
\usepackage{tikz}
\usetikzlibrary{svg.path}

\definecolor{orcidlogocol}{HTML}{A6CE39}
\tikzset{
  orcidlogo/.pic={
    \fill[orcidlogocol] svg{M256,128c0,70.7-57.3,128-128,128C57.3,256,0,198.7,0,128C0,57.3,57.3,0,128,0C198.7,0,256,57.3,256,128z};
    \fill[white] svg{M86.3,186.2H70.9V79.1h15.4v48.4V186.2z}
                 svg{M108.9,79.1h41.6c39.6,0,57,28.3,57,53.6c0,27.5-21.5,53.6-56.8,53.6h-41.8V79.1z M124.3,172.4h24.5c34.9,0,42.9-26.5,42.9-39.7c0-21.5-13.7-39.7-43.7-39.7h-23.7V172.4z}
                 svg{M88.7,56.8c0,5.5-4.5,10.1-10.1,10.1c-5.6,0-10.1-4.6-10.1-10.1c0-5.6,4.5-10.1,10.1-10.1C84.2,46.7,88.7,51.3,88.7,56.8z};
  }
}

\newcommand\orcidicon[1]{\href{https://orcid.org/#1}{\mbox{\scalerel*{
\begin{tikzpicture}[yscale=-1,transform shape]
\pic{orcidlogo};
\end{tikzpicture}
}{|}}}}
\usepackage{amsmath}
\usepackage{amssymb, amsfonts}
\usepackage{babel}
\usepackage[normalem]{ulem}

\begin{document}

\preprint{APS/123-QED}

\title{Quantum kinetic theory of Jeans instability in non-minimal matter-curvature coupling gravity
}

\author{Cláudio Gomes \orcidicon{0000-0001-6022-459X},}\email{claudio.gomes@fc.up.pt}
\address{Faculdade de Ciências e Tecnologia, Universidade dos Açores, Campus de Ponta Delgada, Rua da Mãe de Deus 9500-321 Ponta Delgada, Portugal}
\address{Centro de Física das Universidades do Minho e do Porto, Rua do Campo Alegre s/n, 4169-007 Porto, Portugal}
\address{Institute of Marine Sciences - OKEANOS, University of the Azores, Horta, Portugal}
\author{Kamel Ourabah \orcidicon{0000-0003-0515-6728},}\email{kam.ourabah@gmail.com}
\address{Theoretical Physics Laboratory, Faculty of Physics, University of Bab-Ezzouar, USTHB, Boite Postale 32, El Alia, Algiers 16111, Algeria}
\date{\today}

\begin{abstract}
We present a quantum treatment of the Jeans gravitational instability in the Newtonian limit of the  non-minimal matter-curvature coupling gravity model. By relying on Wigner functions, allowing for the representation of quantum states in a classical phase space, we formulate a quantum kinetic treatment of this problem, generalizing the classical kinetic approach [C. Gomes, \href{https://doi.org/10.1140/epjc/s10052-020-8189-y}{Eur. Phys. J. C \textbf{80}, 633 (2020)}]. This allows us to study the interplay between non-minimal matter-curvature coupling effects, quantum effects, and kinetic (finite-temperature) effects, on the Jeans criterion. We study in detail special cases of the model (general relativity, $f(R)$ theories, pure non-minimal coupling, etc.) and confront the model with the observed stability of Bok globules. 
\end{abstract}

\maketitle

\onecolumngrid
\section{Introduction}

The success of general relativity (GR) is widely acknowledged, as the theory has been confirmed by many observations \cite{Will,bertolamireview} and has been able to predict astrophysical objects like black holes \cite{Barack}. However, despite its predictive success and elegance, GR does not come flawless. In fact, in order to match observations on galactic and cosmological scales, the theory requires two unknown components, namely dark matter (to explain galactic rotation curves and a dynamical mass on galaxy clusters) and dark energy (to explain the late-time accelerated expansion of the Universe).  The lack of detection of these exotic components is the most important challenge faced by GR \cite{mg0,mg1,mg2,mg3}. Besides, the theory also faces other limitations such as issues with unification with high energy physics \cite{mg4,mg5} and the existence of space-time singularities \cite{mg6}. These drawbacks are the main motivations for modified theories of gravity.

There are many possibilities for such modified theories of gravity; some proposals consist in a generalization of the Einstein-Hilbert gravitational action \cite{smg1}, other approaches include minimally or non-minimally coupled scalar fields \cite{smg2,smg3} or additional geometric ingredients \cite{smg4}, while other proposals treat the physical constants as dynamical quantities \cite{smg5,smg6}. Among the simplest modified theories of gravity are the so-called $f(R)$
theories which replace the Ricci scalar by a generic function
thereof in the action functional \cite{f1,f2}. One specific proposal of such theories was initially introduced to deal with the initial conditions problem of the standard Hot Big Bang model, namely the so-called Starobinsky model \cite{starobinskyinflation}. This proposal has many interesting virtues as it is in excellent agreement
with the most recent data from Planck mission \cite{f3}. Besides, it has been shown that requiring $f(R)$ models to be regular at $R = 0$ leads to a behavior
compatible with an effective cosmological constant in a sufficiently curved spacetime which disappears in a flat spacetime \cite{f4}.

Another promising avenue to solve the aforementioned problematic aspects of GR relies on
the extension of $f(R)$ theories with a non-minimal coupling
between matter and curvature \cite{NMC} (hereafter NMC model for short). This approach has a plethora of interesting theoretical and observational implications: It allows for a
mimicking effect of dark matter effects at the scale of galaxies \cite{NMC1} and
galaxy clusters \cite{NMC2}, it accounts for the accelerated expansion of the Universe and provides (through mimicking) a viable unification of dark energy and dark matter \cite{NMC3}, and it is compatible with Planck’s
inflation data \cite{NMC4}, gravitational waves measurements \cite{NMC5},
and the modified virial theorem from the spherically relaxed Abell 586 cluster \cite{NMC6}. Besides, it has been shown \cite{NMC7} that this model yields a correction to
the Newtonian potential, offering therefore interesting possibilities to test this proposal in the weak field regime.

As most interesting phenomena allowing to test modified theories of gravity, and to constraint their parameters, occur in $N$-body systems, one needs to implement statistical methods in these modified theories. The kinetic theory has been recently formulated \cite{NMCB} in the context of the NMC alternative gravity model. It has been shown that an extra force term arises in the Boltzmann equation, due to the geodesic deviation force. Later on, this kinetic treatment has been employed, by one of the present authors, to study the so-called Jeans instability \cite{Claudio}, i.e., the phenomenon responsible for the collapse of a gravitationally bound system, e.g. the interstellar gas, ultimately leading to star formation. The effects of the NMC alternative gravity model on the Jeans criterion for gravitational instability have been studied, offering new possibilities to test the model and to constraint its parameters.

In this paper, we generalize this kinetic treatment to the quantum regime. For that purpose, we formulate a quantum kinetic approach to the NMC generalization of the Schrödinger-Poisson (SP) model [sometimes called Schrödinger-Newton model] (see also \cite{Ferreira1,Ferreira2,Ferreira3} for numerical simulations of NMC extensions of the SP model). The SP model relies on the combination of the Schrödinger equation and the Poisson equation describing the self-potential. Historically, the SP model was first advocated by Diósi \cite{SN1} and Penrose \cite{SN2} as a simple quasi-classical model to
introduce quantum effects in gravitational problems. It describes quantum matter confined by gravitational fields and, as such, finds many applications in astrophysics; it describes (hypothesized) boson stars \cite{BS1, BS2} (which could be a source of ‘exotic’ Laser Interferometer Gravitational-Wave Observatory (LIGO) detections \cite{LIGO}) and it is a central ingredient of scalar field dark matter models \cite{SFDM1,SFDM2,SFDM3,SFDM4}. Interestingly, the same model applies equally well to a variety of other systems,
such as quantum plasmas \cite{pl0,pl1,pl2} and atomic gases in magneto-optical traps \cite{MOT1,MOT2,MOT3}; a feature that may
inspire future laboratory
experiments to test alternative gravity theories using condensed-matter gravity analogs \cite{ga0,ga1,ga2}.
  
By employing the Wigner-Moyal procedure \cite{Wigner,Moyal}, allowing to represent quantum states in a classical phase space, we formulate here a quantum kinetic approach generalizing the classical approach to matter confined by gravitational interactions in the context of the NMC alternative gravity model \cite{Claudio}, and study the NMC effect on the Jeans instability criterion. This allows us to study the rich interplay between NMC effects, quantum effects, and kinetic effects. The rest of the paper progresses in the following fashion: For completeness, we lay out in Sec. \ref{sec:NMC} the theoretical background of the NMC gravity model; we present in Sec. \ref{sec:Wigner} the corresponding Wigner kinetic formulation. In Sec. \ref{sec:jeansinstability}, we establish the dispersion relation for Jeans instability and discuss its limits for negligible (zero-temperature) and dominant kinetic effects. In Sec. \ref{sec:specific}, we analyze in more details specific models and confront the model with data of Bok globules in Sec. \ref{sec:AS}. We conclude in Sec. \ref{sec:conclusion}.

\section{non-minimal matter-curvature coupling model}\label{sec:NMC}

The NMC gravity model is defined through its action functional \cite{NMC}

\begin{equation}\label{S}
S=\int d^{4} x \sqrt{-g}\left[\kappa f_{1}(R)+f_{2}(R) \mathscr{L}\right],
\end{equation}
where $\kappa := c^4/ 16 \pi G$, $\mathscr{L}$ is the Lagrangian density
of matter fields, and $f_1(R)$ and $f_2(R)$ are arbitrary functions of the curvature
scalar $R$. By varying the action (\ref{S}) with respect to the metric $g_{\mu \nu}$, one obtains the metric field equations
\begin{equation}
\Theta G_{\mu \nu}=\frac{1}{2 \kappa} f_{2}(R) T_{\mu \nu}+\Delta_{\mu \nu} \Theta+\frac{1}{2} g_{\mu \nu}\left[f_{1}(R)-R \Theta\right],
\end{equation}
where $G_{\mu \nu}$ is the Einstein tensor and $\Theta:=\left(f_{1}^{\prime}(R)+\frac{f_{2}^{\prime}(R) \mathscr{L}}{\kappa}\right)$, with the primes denoting derivatives with respect to the curvature
scalar $f_{i}^{\prime}(R) \equiv d f_{i}(R) / d R$ and $\Delta_{\mu v} \equiv \nabla_{\mu} \nabla_{v}-g_{\mu v} \square$. Note that for the particular choice $f_1(R)=R$ and $f_2(R)=1$, one recovers GR. The trace of the metric field equations reads as
\begin{equation}
\Theta R-2 f_{1}(R)=\frac{1}{2 \kappa} f_{2}(R) T-3 \square \Theta.
\end{equation}
Hence, the energy-momentum tensor is not covariantly conserved 
\begin{equation}
\nabla_{\mu} T^{\mu v}=\left(g^{\mu v} \mathscr{L}-T^{\mu v}\right) \nabla_{\mu} \ln f_{2}(R).
\end{equation}
This is in fact one of the sticking features of the NMC model. It implies that for a perfect fluid, with $T_{\mu \nu}=( \rho + P) u^{\mu} u^{v}-P g_{\mu v}$, a test particle will not follow geodesic lines, due to
the presence of an extra force term in the geodesics
equation. That is \cite{NMC},
\begin{equation}
\frac{d u^{\alpha}}{d s}+\Gamma_{\mu v}^{\alpha} u^{\mu} u^{v}=f^{\alpha},
\end{equation}
where $u^{\mu}$ is the $4$-velocity of the particle and the extra-force per unit mass $f^{\alpha}$ reads as
\begin{equation}
f^{\alpha}=\frac{1}{\rho+P}\left[\frac{f_{2}^{\prime}(R)}{f_{2}(R)}\left(\mathscr{L}_{m}-P\right) \nabla_{v} R-\nabla_{v} P\right] V^{\alpha v},
\end{equation}
$V^{\alpha \nu}=g^{\alpha \nu}+u^{\alpha} u^{\nu}$ being the projection operator. 
In the case of dust $P=0$ where the Lagrangian has a clear choice $\mathscr{L}=- \rho c^2$, in the Newtonian regime, the geodesic equation reads as \cite{NMC7}
\begin{equation}
\frac{d^{2} x^{i}}{d t^{2}}=\partial_{i}\left[\frac{g_{t t}+1}{2}-\ln \left|f_{2}\right|\right]-\frac{\partial_{i} P}{\rho+P  },
\end{equation}
from which one may define a NMC potential \cite{NMC7} $\Phi_c := \ln \lvert f_2 \rvert$. We note that we shall use units where $c=1$ throughout the rest of the paper, until Sec. \ref{sec:AS}, where we shall recover the International System of Units.

As shown in \cite{Claudio}, for the obvious choice for the metric field 
\begin{equation}
g_{\mu \nu}=\operatorname{diag}(-1-2 \Phi, 1-2 \Psi, 1-2 \Psi, 1-2 \Psi)
\end{equation}
where both $\lvert \Phi \rvert$ and $\lvert \Psi \rvert \ll1$ correspond to the Bardeen
gauge invariant potentials, the $00$-components of the Ricci tensor are $\delta R_{00}=\nabla^{2} \Phi$ and the scalar curvature $\delta R = 2 \nabla^{2}(2 \Psi-\Phi)$. In this case, the two potentials can be decoupled into two
Poisson-like equations \cite{Claudio}

\begin{equation}\label{NMCP}
\left\{\begin{array}{l}
\left(3 \alpha \nabla^{4}-\nabla^{2}\right) \Phi=\left(\alpha \gamma-\frac{\beta}{2}\right) \nabla^{2} \rho-\frac{\gamma}{4} \rho \\
\left(3 \alpha \nabla^{4}-\nabla^{2}\right) \Psi=\left(\frac{\alpha \gamma+\beta}{2}\right) \nabla^{2} \rho-\frac{\gamma}{4} \rho
\end{array}\right.
\end{equation}
where $\rho$ is the mass density (the term $\nabla^2 \rho$ originating
from the non-minimal coupling) and the model parameters are given by $\alpha:=f_1^{''}(0)/f_1'(0)$, $\beta:=f_2'(0)/(\kappa f_1'(0))$ and $\gamma:=f_2(0)/(\kappa f_1'(0))$. Note that by setting $f_2(R)=0$, one recovers $f(R)$ theory while by further setting $f_1(R)=R$ (equivalently, $\alpha=0$), one recovers $\Psi = \Phi$ and the standard Poisson equation is recovered.  

\section{The Wigner equation in the Newtonian limit of the
NMC}\label{sec:Wigner}
In a kinetic approach, one considers the phase space spanned by the space and velocity coordinates $(\mathbf{r},\mathbf{v})$. A state of the given system is characterized by the one-particle distribution function $f(\mathbf{r},\mathbf{v},t)$, such that $f(\mathbf{r},\mathbf{v},t)d^3\mathbf{r}d^3\mathbf{v}$ gives the number of the particles in the volume element $d^3\mathbf{r}$ about the position $\mathbf{r}$ and with velocity in the range $d^3\mathbf{v}$ about $\mathbf{v}$. The space-time evolution of the one-particle distribution function $f$ in the phase space is given by the Boltzmann equation. For a collisionless medium in the presence of a gravitational potential $\Phi$ and the coupling potential $\Phi_c := \ln \lvert f_2 \rvert$ arising from the NMC coupling, the Boltzmann equation reads as \cite{NMCB,Claudio}
\begin{equation}\label{Boltzmann}
\frac{\partial f}{\partial t}+\mathbf{v} \cdot \nabla f-\nabla\left(\Phi+\Phi_{c}\right) \cdot \frac{\partial f}{\partial \mathbf{v}}=0.
\end{equation}
A quantum version of the collisionless Boltzmann equation (\ref{Boltzmann}) can be constructed by employing Wigner functions. for that purpose, we consider a NMC extension of the Schrödinger equation, that is
\begin{equation}\label{Sch}
i \hbar \frac{\partial \psi (\mathbf{r},t)}{\partial t}=-\frac{\hbar^{2}}{2 m} \nabla^2 \psi (\mathbf{r},t) + V \psi (\mathbf{r},t),
\end{equation}
where we have defined a generic potential $V \equiv \Phi + \Phi_c$ as the sum of the gravitational potential $\Phi$ and the nonminimally coupled potential $\Phi_c$. To obtain the quantum counterpart of the Boltzmann equation, we follow the so-called Wigner-Moyal procedure \cite{Wigner,Moyal} (see also \cite{Tito,Our1,Our2}), which allows the representation of quantum states in a classical phase space. To do this, let us first define the Wigner distribution function\footnote{Formally, the Wigner
function is not a \textit{bona fide} distribution and should
be rather regarded as a quasi-distribution, since it can
take negative values. It is nonetheless a very
useful mathematical tool, especially well-suited for
understanding the quantum/classical transition.} associated with the wave-function $\psi$ as follows 

\begin{equation}\label{Wf}
W(\mathbf{r}, \mathbf{p}; t)= \frac{1}{(2 \pi \hbar)^3} \int d \mathbf{y} \exp (i \mathbf{p}. \mathbf{y} / \hbar^3)\psi^{*}(\mathbf{r}+\mathbf{y} / 2, t) \times \psi(\mathbf{r}-\mathbf{y} / 2, t),
\end{equation}
where $\mathbf{p} \equiv m \mathbf{v}$ is the particle momentum. The Wigner function (\ref{Wf}) is simply the Fourier transform of the auto-correlation function corresponding to the wave-function $\psi$. It is normalized here such that
\begin{equation}
\int d \mathbf{p} W(\mathbf{r}, \mathbf{p}; t)=\left|\psi(\mathbf{r}, t)\right|^{2} = \rho(\mathbf{r},t),
\end{equation}
where $\rho(\mathbf{r},t)$ denotes the mass density. Following the Wigner-Moyal procedure, one may write the Schrödinger equation (\ref{Sch}) in the form of a kinetic equation as follows (see for instance \cite{Tito} for detailed calculations)
\begin{equation}\label{WM}
\left(\frac{\partial}{\partial t}+\frac{\mathbf{p}}{m} \frac{\partial}{\partial \mathbf{r}}\right) W-\frac{2 (\Phi + \Phi_c)}{\hbar}  \sin \left(\frac{\hbar}{2} \frac{\overleftarrow{\partial}}{\partial \mathbf{r}} \frac{\overrightarrow{\partial}}{\partial \mathbf{p}}\right) W=0,
\end{equation}
where the sine operator is defined in terms of its Taylor
expansion and the arrows indicate the sense according
to which the operators act; derivatives with respect to
the momentum act forward on the Wigner function while
derivatives with respect to the position act backward on
the potential. Note that in the limit $\hbar \rightarrow 0$, Eq. (\ref{WM}) reduces to the NMC Boltzmann equation (\ref{Boltzmann}).
This formal limit corresponds to the case of a slowly varying potential, changing significantly only over a length-scale much larger
than the de Broglie wavelength, such that $\sin \Lambda \simeq \Lambda$.

In what follows, we shall use the Wigner kinetic equation (\ref{WM}) together with the NMC Poisson-like equations (\ref{NMCP}) to address the collective behavior of a self-gravitating medium and the associated Jeans instability. This quantum kinetic treatment allows to investigate the interplay between NMC effects, quantum effects, and kinetic effects, covering therefore a wide range of situations. 

\section{Jeans Instability} \label{sec:jeansinstability}
Using the Wigner equation (\ref{WM}), we follow here the standard procedure yielding the dispersion relation (see for instance \cite{Claudio}). We restrict ourselves to the case of linear perturbations and write the relevant quantities as a small perturbation, in the form of plane waves, around homogeneous and stationary quantities (given by $W_0(\mathbf{p})$, $\Phi_0=0$, and $\Psi_0=0$). That is 
\begin{equation}
\begin{aligned}
W(\textbf{r}, \textbf{p} ; t) & \equiv W_{0}(\textbf{p})+\tilde{W} \exp [i(\textbf{k}. \textbf{r}-\omega t)] \\
\Phi(\textbf{r}, t) & \equiv \Phi_0 + \tilde{\Phi} \exp [i(\textbf{k}.\textbf{r} -\omega t)], \\
\Psi(\textbf{r}, t) & \equiv \Psi_0+ \tilde{\Psi} \exp [i(\textbf{k}.\textbf{r} -\omega t)],
\end{aligned}
\end{equation}

By linearizing the Wigner equation (\ref{WM}) and making use of the so-called Jeans swindle, i.e., by
considering that the potentials are sourced only by
the density perturbations and not by the density background, we obtain the following dispersion relation 
\begin{equation}\label{IDR}
1+\frac{(\alpha \gamma-\beta / 2) k^{2}+\gamma / 4}{k^{2}+3 \alpha k^{4}} \frac{m}{\hbar} \int_{- \infty}^{\infty} \frac{ f_0(p+\hbar k/2)-f_0(p-\hbar k/2)}{p k/m-\omega} d {p}=0,
\end{equation}
where we have defined $f_0(p)$ as the one-dimensional projected (marginal) distribution along the axis parallel to the wave-vector $\mathbf{k}$. That is,
\begin{equation}
f_0(p)=\iint W_{0}\left(p, \mathbf{p_{\perp}} \right) {d \mathbf{p_{\perp}}},
\end{equation}
{where $p$ and $\mathbf{p_{\perp}}$ stand for the parallel and perpendicular components of the momentum respectively, i.e.,}

\begin{equation}
\mathbf{p}=p \frac{\mathbf{k}}{k}+\mathbf{p}_{\perp}.
\end{equation}

It may be noted that, in the limit $\hbar \rightarrow 0$, the dispersion relation (\ref{IDR}) reduces to the classical one  \cite{Claudio}
\begin{equation}
1+\frac{(\alpha \gamma-\beta / 2) k^{2}+\gamma / 4}{k^{2}+3 \alpha k^{4}} \int_{-\infty}^{\infty} \frac{ {\partial f_0}/{\partial {v}}}{{v}-\omega /k} d {v}=0,
\end{equation}
with $v$ being the component of the velocity parallel to the wave-vector $\mathbf{k}$. The dispersion relation (\ref{IDR}) is very general and contains both quantum and kinetic effects. It may be interesting to study particular limits where kinetic effects are negligible or dominant.

\subsection{The zero temperature limit}
\label{sec:zerotemperature}

It may be interesting to analyze the zero-temperature limit (negligible kinetic effects), where the equilibrium distribution $f_0$ reduces to a Dirac delta, i.e.,
\begin{equation}\label{delta}
f \equiv \rho_0 \delta (p).
\end{equation}
In this case, the integral dispersion relation (\ref{IDR}) reads as

\begin{equation}\label{dr21}
1- \frac{(\alpha \gamma-\beta / 2) k^{2}+\gamma / 4}{k^{2}+3 \alpha k^{4}}  \frac{m \rho_0}{\hbar} \left [  \frac{1}{\omega+\hbar k^2 /2m}-\frac{1}{\omega-\hbar k^2 /2m} \right ]=0.
\end{equation}
After some simple algebraic manipulations, Eq. (\ref{dr21}) can be written as

\begin{equation}\label{omega}
\omega^2=-\frac{(\alpha \gamma-\beta / 2) k^{2}+\gamma / 4}{1+3 \alpha k^{2}} \rho_0   + \frac{\hbar^2 k^4}{4m^2}.
\end{equation}
The latter equation shows that quantum effects (i.e., quantum pressure) act against gravitational instability, preventing gravitational collapse to occur. This may be made more transparent by observing that, in the limit corresponding to GR, i.e., $\alpha=\beta=0$ and $\gamma=1/\kappa$, Eq. (\ref{omega}) reduces to the well-known dispersion relation (see for instance \cite{Her})
\begin{equation}
\omega^2=- \Omega_J^2 + \frac{\hbar^2 k^4}{4m^2},
\end{equation}
where $\Omega_J=\sqrt{4 \pi G \rho_0}$ is the so-called Jeans frequency.

Note that for $\omega^2>0$, the frequency is real and the perturbation behaves as $\operatorname{e}^{-i \omega t}$, i.e., harmonic waves, while for $\omega^2<0$, the frequency
is imaginary ($\omega = i \gamma$) and the perturbation behaves as $\operatorname{e}^{\gamma t}$, i.e., it evolves exponentially
with time with a rate $\gamma$ (with $\gamma = \pm \sqrt{- \omega^2}$). There is a growing mode ($\gamma >0$) and a decaying
mode ($\gamma <0$). The growing mode is responsible for Jeans instability. The sign of $\omega^2$ in (\ref{omega}) determines the threshold value of $k$ separating between an oscillatory regime ($\omega$ real) and exponential growth or instability ($\omega$ imaginary). Setting $\omega=0$, Eq. (\ref{omega}) can be solved for $k^2$. By defining $\xi=m^2 \rho_0/\hbar^2$, the general solution, provided that $\alpha\neq 0$, can be written as

 \begin{equation}\label{k2t0}
k^2= -\frac{1}{9\alpha}\left[1+\Delta+\frac{1-18 \alpha  \xi  (\beta -2 \alpha  \gamma )}{\Delta}\right]~,
 \end{equation}
 where $\Delta^3=1-\alpha\xi\left(27\beta+\frac{135}{2}\alpha\gamma\right)+\sqrt{\left(1-\alpha\xi\left(27\beta+\frac{135}{2}\alpha\gamma\right)\right)^2-\left(1-18\alpha\xi (\beta-2\alpha\gamma) \right)^3}$. This is the critical wave number separating between oscillatory and unstable modes. Perturbations characterized by a wave number smaller than the critical wave number (\ref{k2t0}) are unstable.

{Before closing this subsection, it is instructive to note that, although using the language of the kinetic theory, kinetic effects have not been considered so far, since the equilibrium distribution $f_0(p)$ has been identified with a Dirac delta (\ref{delta}). This limit can be studied in a more straightforward way by relying on the quantum hydrodynamic model (QHM). To show that, let us go back to the Schrödinger equation (\ref{Sch}) and write the wave-function in polar form, that is
\begin{equation}\label{psi}
\psi(\mathbf{r},t)= \sqrt{\rho(\mathbf{r},t)}  \exp \left(i S(\mathbf{r},t) / \hbar\right),
\end{equation}}

{where $\rho = \left|\psi\right|^{2}$ is the density and
\begin{equation}
S(\mathbf{r},t)=-i \frac{\hbar}{2} \ln \left(\frac{\psi}{\psi^{*}}\right)
\end{equation}
is the phase of the wave-function. Following the \textit{Madelung-Bohm procedure} \cite{Madelung,Bohm}, one may define a fluid velocity field such that
\begin{equation}\label{Mad2}
\quad \mathbf{u} =\frac{1}{m} \nabla S,
\end{equation}} 
{which ensures that the current is given by
\begin{equation}
\mathbf{j} =\frac{\hbar}{2 m i}\left[\psi^{*}(\nabla \psi)-\psi\left(\nabla \psi^{*}\right)\right]= \rho \mathbf{u} .
\end{equation}} 
{Substituting the wave-function (\ref{psi}) into the Schrödinger equation (\ref{Sch}) and splitting apart the real and imaginary parts, one obtains
\begin{equation}\label{HQ}
\begin{aligned}
\frac{\partial \rho}{\partial t}+\nabla\left(\rho \mathbf{u}\right) &=0, \\
m \left(\frac{\partial \mathbf{u}}{\partial t}+\mathbf{u} \nabla \mathbf{u} \right) &=-{\nabla \Phi} -{\nabla \Phi_c}  - \nabla Q,
\end{aligned}
\end{equation}}
{where 
\begin{equation}
Q \equiv  -\frac{\hbar^{2}}{2 m}  \left(\frac{\nabla^{2}\sqrt{\rho}}{\sqrt{\rho}}\right) =-\frac{\hbar^{2}}{4 m}\left[\frac{\nabla^2 \rho}{\rho}-\frac{1}{2} \frac{(\nabla \rho)^{2}}{\rho^{2}}\right]
\end{equation}}
{is the so-called \textit{quantum Bohm potential}.} {The first equation in Eq. (\ref{HQ}) is the \textit{continuity equation} while the second one is the quantum \textit{Euler equation}. The set of equations (\ref{HQ}) may be employed following standard lines (linearization and decomposition in Fourier modes) to obtain the dispersion relation (\ref{omega}). Through these lines, it is more explicit that the term $\hbar^2 k^4/4m^2$ in Eq. (\ref{omega}) arises from the quantum pressure force $-\nabla Q$ which acts against gravity, stabilizing the self-gravitating instability for small wavelengths.}

\subsection{Kinetic effects}
To analyze the effect of the NMC gravity model in astrophysical scenarios, it may be necessary to account for purely kinetic effects, i.e., finite-temperature corrections to the distribution function. For that purpose, we go back to the general dispersion relation (\ref{IDR}), by considering a general distribution $f_0(p)$ other than the Dirac delta (\ref{delta}).
We assume an \textit{even} distribution $f_0(p)$, that is
\begin{equation}
\int_{- \infty}^{\infty} p f_{0}(p) \mathrm{d} p=0,
\end{equation}
which is characteristic for equilibrium and nearly equilibrium situations, and consider small quantum effects ($\hbar k/2 \ll p$). In this case, one may Taylor expand the functions $f_0({p+ \hbar k/2})$ and $f_0({p- \hbar k/2})$ and write the integral in the dispersion relation (\ref{IDR}) as

\begin{equation}
\int_{- \infty}^{\infty} \frac{f_0({p+ \hbar k/2})-f_0({p- \hbar k/2})}{p k/m- \omega} d {p} \approx \hbar k \int_{- \infty}^{\infty} \frac{f_0'(p) dp}{p k/m- \omega} + \frac{\hbar^3 k^3}{24} \int_{- \infty}^{\infty} \frac{f_0''(p) dp}{p k/m- \omega}.
\end{equation}
We consider the limit of long wavelengths ($p \gg m \omega / k$), so that the singularity in the denominator is avoided. By performing a Taylor expansion of $(p-m \omega / k)^{-1}$ and keeping only terms with small values of $k$ (long wavelength), one obtains the following dispersion relation
\begin{equation}
\label{DRT}
\omega^{2} \simeq -\frac{(\alpha \gamma-\beta / 2) k^{2}+\gamma / 4}{1+3 \alpha k^{2}} \rho_0 +3 \frac{k_B T}{m} k^{2}+\frac{\hbar^{2} k^{4}}{4 m^{2}},
\end{equation}
where we have defined the temperature in the kinetic sense, i.e.,
\begin{equation}
 {k_B T} \equiv \frac{\int  p^{2}/m f_{0}(p) \mathrm{d} p}{\int f_{0}(p) \mathrm{d} p},
\end{equation}
$k_B$ being the Boltzmann constant. The dispersion relation (\ref{DRT}) generalizes Eq. (\ref{omega}) by incorporating the effect of a non-vanishing velocity dispersion (thermal corrections). It remains valid as long as quantum corrections are small. It shows that, in addition to the effect of quantum pressure, thermal corrections also act against the instability, preventing the gravitational collapse from occurring. In this case, the solutions for Eq. (\ref{DRT}) when $\omega=0$ (separating between oscillatory and unstable modes) are more elaborated, and can be cast as

\begin{equation}
    k^2=-\frac{1}{9\alpha}\left[(1+36\alpha \xi \epsilon)-\frac{A}{\left(B+\sqrt{B^2+A^3}\right)^{1/3}}+\left(B+\sqrt{B^2+A^3}\right)^{1/3}\right]~,
\end{equation}
where we have defined $\xi:=\frac{m^2 \rho_0}{\hbar^2}$, as before, and $\epsilon:=\frac{k_B T}{m\rho_0}$, $A:=18\alpha\xi\left( (\beta-2\alpha\gamma)+6\epsilon\right)-1-72\xi\alpha\epsilon\xi-1296\xi^2\alpha ^2\epsilon^2$ and $B:=1-\alpha \xi (27\beta+\frac{135}{2}\alpha\gamma)-54\xi\epsilon\alpha(1+18\xi\alpha(\beta-2\alpha\gamma))-1944\xi^2\alpha^2\epsilon^2+46656\xi^3\alpha^3\epsilon^3$. Note that when the nonzero temperature corrections are turned off, we get a smooth transition to the previous scenario (subsection \ref{sec:zerotemperature}), namely $(B+\sqrt{B^2+A^3})=\Delta^3$, and $A\to -1+18 \alpha  \xi  (\beta -2 \alpha  \gamma )$. Note also that in the limit $\hbar \to 0$, one recovers the results of \cite{Claudio} (the classical kinetic regime).

\section{Specific gravity models}\label{sec:specific}

In this section, we shall analyze some relevant particular situations, such as $f(R)$, pure non-minimal coupling, and $2\alpha \gamma = \beta$ models, likewise Ref. \cite{Claudio} in the classical context of Jeans instabilities for this alternative gravity model. These models allow for understanding the differences between quantum corrections with and without nonzero temperature kinetic effects.

 \subsubsection{f(R) theories}
 
\begin{figure}[h!]
     \centering
     \includegraphics[width=0.5\textwidth]{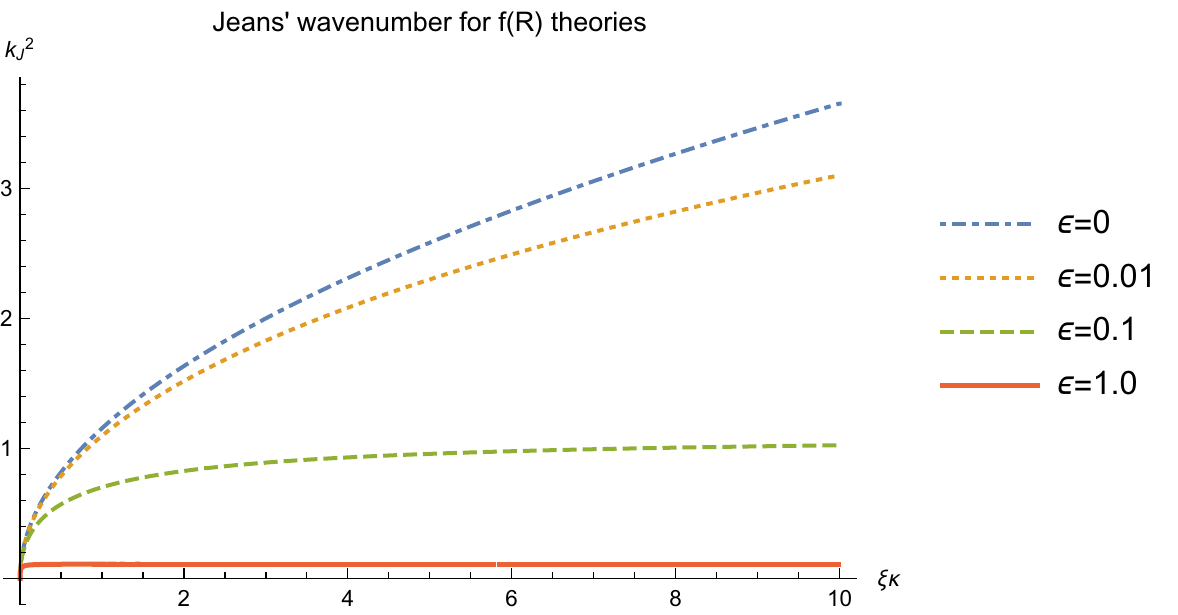}
     \caption{Behavior of the squared modified Jeans wavenumber in f(R) theories as a function of the system parameter $\xi\kappa$ in the limit $\alpha \to -\infty$ for different values of $\epsilon$.}
     \label{fig:alphafrt0}
 \end{figure}

 In this case, we have $\beta=0$ and $\gamma=1/\kappa$, which for $\alpha \neq 0$ in the zero temperature limit yields:
 
 \begin{equation}
  k^2=-\frac{1}{9 \alpha }\left[1+\Delta_1+\frac{\left(1+36\xi\alpha^2/\kappa\right)}{\Delta_1}\right]~,
 \end{equation}
 where $\Delta_1^3=1-\frac{135\xi\alpha^2/\kappa}{2}+\sqrt{\left(1-\frac{135\xi\alpha^2/\kappa}{2}\right)^2-\left(1+36\xi\alpha^2/\kappa\right)^3}$. On the other hand, for $\alpha=0$ which is the case of GR, we have solved Eq. (\ref{omega}) and we get a constant solution $k^2=\sqrt{\frac{\xi }{\kappa}}$. If $\alpha>0$, then $k^2<0$ and imaginary solutions for the modified Jeans mass are found. On the other hand, $\alpha<0$ yields positive solutions, however, we should recall that the condition $f_1''>0$ is demanded for avoiding Dolgov-Kawasacki instabilities in $f(R)$ theories, which implies that the denominator of the definition of the $\alpha$ parameter should be negative, $f_1'(0)<0$. In particular, if we take the limit $\alpha \to -\infty$, we find the behavior of $\tilde{k}_J^2$ as a function of $\xi/\kappa$, which signals the characteristics of the system under analysis, is as shown in Fig. (\ref{fig:alphafrt0}).
 
 When kinetic corrections are included, for $\alpha=0$, we have:
 \begin{equation}
     k^2_{\pm}=-6\epsilon\xi \pm \sqrt{(6\epsilon\xi)^2+\xi/\kappa}~,
 \end{equation}
 where the physical solutions correspond to $k^2_+:=-6\epsilon + \sqrt{(6\epsilon)^2+\xi/\kappa}$, provided that $\xi/\kappa > 0$. For $\alpha \neq 0$, we have:
 \begin{equation}
     k^2=-\frac{1}{9\alpha}\left[1+36\alpha\xi\epsilon-\frac{\bar{A}}{\left(\bar{B}+\sqrt{\bar{B}^2+\bar{A}^3}\right)^{1/3}}+\left(\bar{B}+\sqrt{\bar{B}^2+\bar{A}^3}\right)^{1/3}\right]~,
 \end{equation}
 where we defined $\bar{A}:=-1-36\frac{\alpha^2}{\kappa}\xi+36\alpha\xi\epsilon-1296\xi^2\alpha^2\epsilon^2$ and $\bar{B}=1-\frac{135}{2}\frac{\alpha^2}{\kappa}\xi-54\alpha\xi\epsilon+1944\frac{\alpha^3}{\kappa}\xi^2\epsilon-1944\alpha^2\xi^2 \epsilon^2+46656\alpha^3\xi^3\epsilon^3$. In limit of $\alpha \to - \infty$, we get:
 \begin{equation}
    k^2= \frac{2}{3} \left(\sqrt[3]{-216 \text{$(\xi \kappa)$}^3 \epsilon ^3+\sqrt{-\text{$(\xi \kappa)$}^3 \left(27 \xi \kappa \epsilon ^2+1\right)}-9 \text{$(\xi \kappa)$}^2 \epsilon }+\frac{\text{$(\xi \kappa)$} \left(36 \text{$\xi \kappa$} \epsilon ^2+1\right)}{\sqrt[3]{-216 \text{$(\xi \kappa)$}^3 \epsilon ^3+\sqrt{-\text{$(\xi \kappa)$}^3 \left(27 \xi \kappa \epsilon ^2+1\right)}-9 \text{$(\xi \kappa)$}^2 \epsilon }}-6 \xi \kappa \epsilon \right)
 \end{equation}

 \subsubsection{Pure non-minimal matter-curvature coupling}
 
 This case is characterized by setting $\alpha=0$. In the absence of kinetic corrections, the solution for the Jeans wavenumber is given by:
 
 \begin{equation}
     k^2_{\pm}=-\beta\xi   \pm \sqrt{\gamma\xi +\beta ^2\xi^2 }~,
 \end{equation}
however, only $k^2_+$ yields a physical solution as it provides real solutions for $k$ provided $\gamma>0$, as we want to preserve a positive gravitational coupling during a gravitational collapse scenario.
 
 When kinetic effects are added, the solution becomes:
 \begin{equation}
     k^2_{\pm}=-\beta \xi-6\epsilon\xi \pm \sqrt{\gamma  \xi+\left(6\epsilon\xi+\beta  \xi\right)^2}
 \end{equation}
 
 The positive branch, $k^2_+$, yields positive solutions provided that $\gamma >0$.
 
\begin{figure}[h!]%
    \centering
    \subfloat[\centering Jeans' wavenumber as a function of $\beta\xi$ for fixed values of $\gamma\xi$ due to quantum corrections.]{{\includegraphics[width=8cm]{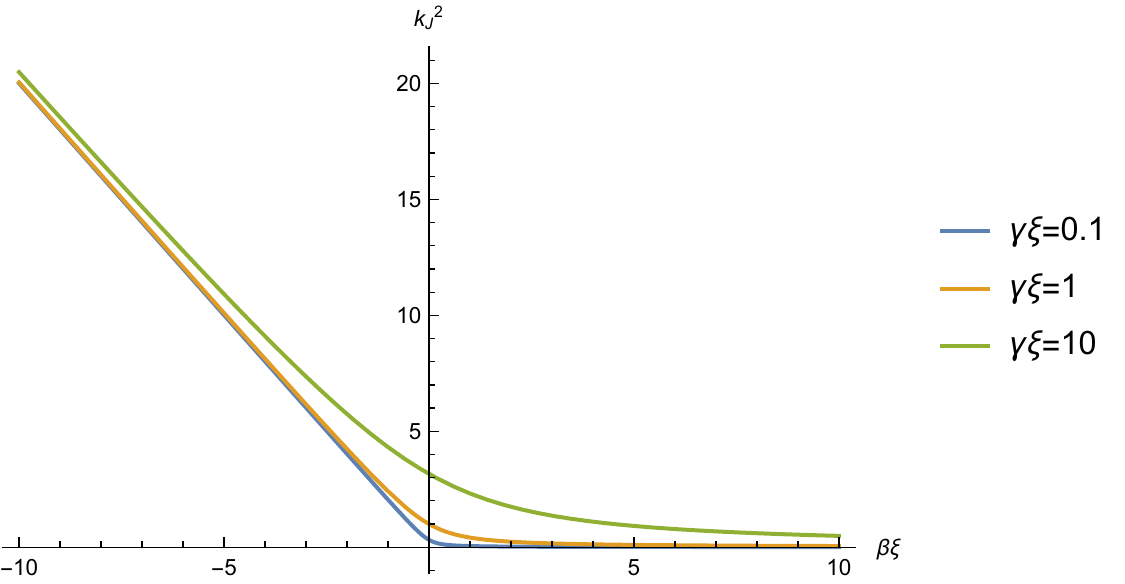} }}%
    \qquad
    \subfloat[\centering Jeans' wavenumber as a function of $\beta\xi$ for fixed values of $\epsilon\xi$ and the fixed value of $\gamma\xi=1$ due to quantum and thermal corrections.]{{\includegraphics[width=8cm]{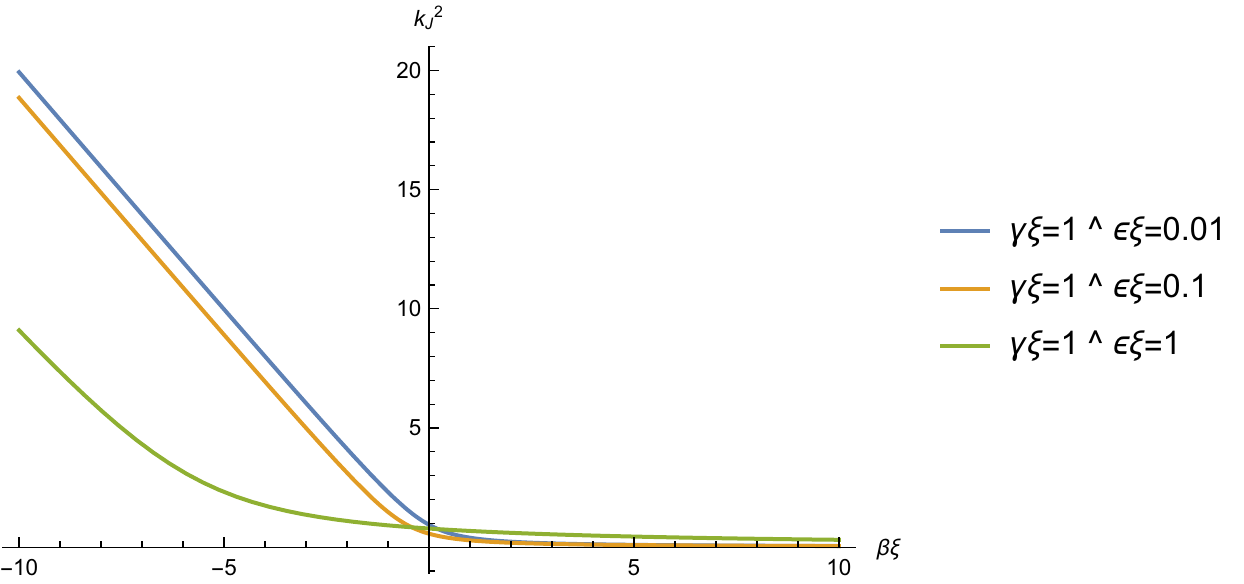} }}%
    \caption{Behavior of the squared modified Jeans' wavelength with only quantum correction, or with quantum and kinetic effects for a pure non-minimal matter-curvature coupling model.}%
    \label{fig:example}%
\end{figure}

 \subsubsection{$2\alpha \gamma = \beta$}
 
This is a special case where a given combination of parameters allows for a simpler analysis. The solutions for the Jeans' wavenumber are given by:
 \begin{equation}
     k^2=-\frac{1}{9 \alpha }\left[1+\Delta_2+\frac{1}{  \Delta_2}\right]
 \end{equation}
where $\Delta_2^3=1-\frac{243 \alpha ^2 \xi \gamma  }{2}+\sqrt{-243 \xi \alpha ^2 \gamma+\left(\frac{243}{2}\right)^2 \xi^2\alpha^4\gamma^2}$.

To ensure real solutions for $k^2$, a further condition can be found: $\alpha^2\gamma=\frac{\alpha \beta}{2}>\frac{4h^2}{243m \rho_0}$. Moreover, $\Delta_2$ needs to be negative to give positive solutions for $k^2$.

When kinetic effects are added, the solution becomes:
\begin{equation}
  k^2=-\frac{1}{9\alpha}\left[1+36\alpha\epsilon\xi-\frac{A_2}{\left(B_2+\sqrt{B_2^2+A_2^3}\right)^{1/3}}+\left(B_2+\sqrt{B_2^2+A_2^3}\right)^{1/3}\right]  
\end{equation}
where $A_2:=108\alpha\epsilon\xi-(1+36\alpha\epsilon\xi)^2$ and $B_2:=1-\frac{243}{2}\alpha^2\gamma\xi-54\alpha\epsilon\xi-1944\alpha^2\epsilon^2\xi^2+46656\alpha^3\epsilon^3\xi^3$. However, as it can be numerically found, there are no real solutions for $k$, either with quantum corrections alone, or with kinetic effects, hence this scenario does not provide a good physical description of a gravitationally collapsing system.

\section{Astrophysical Systems}\label{sec:AS}

One way to assess the physical viability of the previous solutions is to test with regions in the Universe that can experiment star formation. One of such examples is in Bok globules, which are nearby isolated clouds of interstellar gas and dust with simple shapes. They have characteristic temperatures of the order of 10K, and masses of $10 M_{\odot}$ which are close to their corresponding Jeans' masses. 

{The Jeans' mass $M_J$ is defined as the mass inside a sphere of diameter $2\pi/k_J$, where the Jeans' wavenumber reads $k_J^2=\frac{4\pi G \rho_0}{c_s^2}$, being $c_s$ the sound speed. Thus, these astrophysical systems are perfect candidates to test alternative theories of gravity as well as infer whether kinetic and quantum corrections are observable.

Moreover, for the NMC alternative theories of gravity, we have a modified Jeans' mass:
\begin{equation}
\tilde{M}_J=\left(\frac{k_J^2}{\tilde{k}^2_J}\right)^{3/2}M_J~,
\end{equation}
where $\tilde{k}_J$ is the modified Jeans' mass and corresponds to the solutions we have obtained in the previous section, $\tilde{k}_J:=k(\omega=0)$.

Noting that $\rho_0=n_{H_2}\mu m_p$, being $n_{H_2}$ the particle number density and $\mu=2.33$ the mean molecular weight for Bok globules \cite{Kandori}, and $m_p$ the proton mass, we can find that the typical parameters for this system are $\epsilon=\mathcal{O}(10^{20})$ and $\xi=\mathcal{O}(10^{-1})$. This means that the kinetic corrections, $\epsilon$, are several orders of magnitude higher than the quantum ones, $\xi^{-1}$, for Bok globules, hence we can neglect the latter ones. Therefore, these considerations together with the phenomenological factor $\tilde{M}_J=\left(\frac{2}{5}\right)M_J$ found in Ref. \cite{Claudio} as a sufficient condition for matching with the Bok globules stability observations \cite{Kandori}, we can find saturating bounds for the parameters of the gravity models parameters, namely $\alpha, ~\beta, ~\gamma$. That is, for alternative theories of gravity, the following equation, provided the observational saturation bound, has to be solved:
\begin{equation}
\label{eqn:JeansBok}
    \left(\frac{k_J^2}{\tilde{k}^2_J}\right)^{3/2}=\frac{2}{5}~.
\end{equation}

Let us now consider the standard scenario of General Relativity and assess the modifications that appear in the NMC gravity model for the three subclasses analyzed in Sec. \ref{sec:specific}. To this end, we shall resort to power-law functions for both $f_1(R)$ and $f_2(R)$ as it has been shown to correctly address cosmological and astrophysical problems such as dark matter, dark energy, inflation, gravitational waves or the cosmic virial theorem \cite{NMC1,NMC2,NMC3,NMC4,NMC5,NMC6}.

\subsection{General Relativity}

GR together with kinetic corrections yields a modified Jeans wavenumber of the form:
\begin{equation}
    k^2=\frac{1}{12\epsilon \kappa}~,
\end{equation}
and as expected these corrections counteract the gravitational instability, hence we get modified Jeans' masses, $M_J^{(T)}$, which are higher than the observed masses, as depicted in Table \ref{tab:globules}. This results in expecting that kinetic corrections counteract the gravitational collapse, hence providing stability. On the other hand, there are regions which are observed being unstable, hence opening an avenue for allowing alternative theories of gravity, whose additional terms can ensure a suitable gravitational collapse in some scenarios.

We further note that the difference between the computed Jeans mass in General Relativity reported by Ref. \cite{Vainio} and our values lies in the fact that we performed a long wavelength expansion, which resulted in a multiplying factor 3 in the term $k_B T$ of the dispersion relation. This occurs either in a quantum kinetic regime as in our problem \cite{ga1}, or in a pure classical kinetic approach by performing such expansion. Nevertheless, we shall use the values of Ref. \cite{Vainio} as a direct comparison to ours in the scenarios of General Relativity and f(R) theories.

\begin{table}[h!]
\scriptsize
\begin{tabular}{|l|c|c|c|c|c|c|c|} \hline
Bok Globule &$T_0\text{[K]}$ & $n_{H_2}\text{[cm$^{-3}$]}$ & $M [M_{\odot}]$  & $M_J [M_{\odot}]$ & $M_J^{(T)} [M_{\odot}]$ & Stability \\ \hline
CB 87	& 11.4 &	$(1.7\pm 0.2)\times 10^4$	& $2.73\pm 0.24$ & 9.6 & $49,9$ & stable \\ \hline
CB 110 & 21.8 & $(1.5\pm 0.6)\times 10^5$ & $7.21\pm 1.64$ & 8.5  & $44.2$ & unstable \\ \hline
CB 131 & 25.1 & $(2.5\pm 1.3)\times 10^5$ & $7.83\pm 2.35$ & 8.1  & $42.1$ & unstable \\ \hline
CB 134 & 13.2 & $(7.5\pm 3.3)\times 10^5$ & $1.91\pm 0.52$ & 1.8  & $9.4$ & unstable \\ \hline
CB 161 & 12.5 & $(7.0\pm 1.6)\times 10^4$ & $2.79\pm 0.72$ & 5.4   & $28.1$ & unstable \\ \hline
CB 184 & 15.5 & $(3.0\pm 0.4)\times 10^4$ & $4.70\pm 1.76$ & 11.4  & $59.3$ & unstable \\ \hline
CB 188 & 19.0 & $(1.2\pm 0.2)\times 10^5$ & $7.19\pm 2.28$ & 7.7 & $40.0$  & unstable \\ \hline
FeSt 1-457 & 10.9 & $(6.5\pm 1.7)\times 10^5$ & $1.12\pm 0.23$ & 1.4  & $7.3$ & unstable \\ \hline
Lynds 495 & 12.6 & $(4.8\pm 1.4)\times 10^4$ & $2.95\pm 0.77$ & 6.6  & $34.3$ & unstable \\ \hline
Lynds 498 & 11.0 & $(4.3\pm 0.5)\times 10^4$ & $1.42\pm 0.16$ & 5.7 & $29.6$  & stable \\ \hline
Coalsack & 15 & $(5.4\pm 1.4)\times 10^4$ & $4.50$ & 8.1 & $42.1$ & stable \\ \hline
\end{tabular}
\caption{Kinetic temperature, particle number density, mass, Jeans' mass and observed stability from several Bok globules \cite{Kandori,Vainio}, with the General Relativity predictions when kinetic effects are included.}
\label{tab:globules}
\end{table}

\subsection{f(R) theories}

For $f(R)$ theories, the solution reads:
\begin{equation}
   k^2_{\pm}=\frac{\alpha/\kappa   -3 \epsilon \pm \sqrt{(\alpha/\kappa) ^2 +3 \alpha/\kappa  \epsilon +9 \epsilon ^2}}{18 \alpha  \epsilon } ~,
\end{equation}
being $k^2_+$ the only viable physical solution.

By solving Eq. (\ref{eqn:JeansBok}), we get two different solutions for $\alpha$ for each Bok globule, as shown in Table \ref{tab:globulesFR}.

\begin{table}[h!]
\scriptsize
\begin{tabular}{|l|c|c|c|c|c|c|c|} \hline
Bok Globule &$T_0\text{[K]}$ & $n_{H_2}\text{[cm$^{-3}$]}$ & $M [M_{\odot}]$  & $M_J [M_{\odot}]$ & $\alpha_1 [m^2]$ & $\alpha_2 [m^2]$ \\ \hline
CB 87	& 11.4 &	$(1.7\pm 0.2)\times 10^4$	& $2.73\pm 0.24$ & 9.6 & $-3.89\times 10 ^{59}$ & $1.20\times 10^{13}$ \\ \hline
CB 110 & 21.8 & $(1.5\pm 0.6)\times 10^5$ & $7.21\pm 1.64$ & 8.5  & $-1.82\times 10 ^{58}$ &  $-1.52\times 10^{13}$ \\ \hline
CB 131 & 25.1 & $(2.5\pm 1.3)\times 10^5$ & $7.83\pm 2.35$ & 8.1  & $-8.70\times 10^{57}$ & $3.62\times 10^{12}$ \\ \hline
CB 134 & 13.2 & $(7.5\pm 3.3)\times 10^5$ & $1.91\pm 0.52$ & 1.8  & $-2.68\times 10^56$ & $6.40\times 10^{11}$ \\ \hline
CB 161 & 12.5 & $(7.0\pm 1.6)\times 10^4$ & $2.79\pm 0.72$ & 5.4   & $-2.75\times 10^{58}$ & $2.78\times 10^{12}$ \\ \hline
CB 184 & 15.5 & $(3.0\pm 0.4)\times 10^4$ & $4.70\pm 1.76$ & 11.4  & $-2.31\times 10^{59}$ & $3.69\times 10^{13}$ \\ \hline
CB 188 & 19.0 & $(1.2\pm 0.2)\times 10^5$ & $7.19\pm 2.28$ & 7.7 & $-2.18\times 10^{58}$  & none \\ \hline
FeSt 1-457 & 10.9 & $(6.5\pm 1.7)\times 10^5$ & $1.12\pm 0.23$ & 1.4  & $-2.44\times 10^{56}$ & $6.67\times 10^{11}$ \\ \hline
Lynds 495 & 12.6 & $(4.8\pm 1.4)\times 10^4$ & $2.95\pm 0.77$ & 6.6  & $-5.95\times 10^{58}$ & $3.46\times 10^{13}$ \\ \hline
Lynds 498 & 11.0 & $(4.3\pm 0.5)\times 10^4$ & $1.42\pm 0.16$ & 5.7 & $-5.65\times 10^{58}$  & $3.62\times 10^{13}$ \\ \hline
Coalsack & 15 & $(5.4\pm 1.4)\times 10^4$ & $4.50$ & 8.1 & $-6.64\times 10^{58}$ & $3.86\times 10^{13}$ \\ \hline
\end{tabular}
\caption{Kinetic temperature, particle number density, mass, Jeans' mass and observed stability from several Bok globules \cite{Kandori,Vainio}, and the saturation bounds found for $\alpha$ in f(R) theories.}
\label{tab:globulesFR}
\end{table}

We can explore a particular model of this scenario, namely the Starobinsky model $f_1(R)=R+aR^2$ \cite{starobinskyinflation}, for which $\alpha=2a$, and the results obtained follows from the above Table. In particular, we can note that in the Starobinky model $2a\approx 3,84\times 10^{13}$ which is remarkably similar to the values found for $\alpha_2$.

Furthermore, for $f_1(R)=R+aR^n$, for $n>1$, we have $\alpha=0$, and we expect to have the same results as in the General Relativity case. We note that this occurs since we have developed the Jeans analysis for a low field expansion, as it is standard. However, if the background spacetime had more curvature/nonlinear effects, the corresponding, and much more elaborated, form of $\alpha$ would differ from its GR counterpart.

\subsection{Pure NMC}

The pure non-minimal matter-curvature coupling gravity model yields the following solution for the Jeans wavenumber:
\begin{equation}
    k^2=\frac{\gamma}{2\beta + 12 \epsilon}~,
\end{equation}
which in its turn, by solving equation Eq. (\ref{eqn:JeansBok}), we get the following relation:
\begin{equation}
\label{eqn:relationpurenmc}
    \gamma =  50^{1/3} (\beta +6 \epsilon ) k_J^2~.
\end{equation}

Let us consider a linear pure non-minimal coupling function, namely $f_2(R)=1+b R$, which has been recently constrained to have a coupling $|b| < 2\times 10^{-12}m^2$ \cite{nuclearconstraints}. This model has $\gamma=1/\kappa$ and $\beta=b/\kappa$, hence we can estimate the modified Jeans' mass as $\tilde{M}_J=5.2M_J$. Therefore, we can conclude that this linear functions is not suitable for Bok globules.

We can further consider a general function $f_2(R)=1+b R^n$, for $n>1$, for which $\gamma=1/\kappa$ and $\beta=0$, which also fails for explaining Bok globules. Therefore, the non-minimal coupling alone, given by a power-law function, is not sufficient to explain the observed (in)stability of Bok globules.

\subsection{$2\alpha\gamma=\beta$}

In this particular case of gravity theories, we get:
\begin{equation}
    k^2_{\pm}=\frac{-\epsilon \pm \sqrt{\alpha \gamma \epsilon+\epsilon^2}}{6\alpha\epsilon}~,
\end{equation}
where only the positive sign solution is physically meaningful, thus yielding a relation between $\alpha$ and $\gamma$, by solving Eq. (\ref{eqn:JeansBok}), which is given by:
\begin{equation}
    \gamma = \frac{3\times\sqrt[3]{10}\left(30 \sqrt[3]{2} \alpha  \epsilon  k_J^4+\sqrt[3]{5}\epsilon k_J^2\left(4+675 \alpha ^3 k_J^2+\left|4-675 \alpha ^3 k_J^6\right|\right){}^{\frac{2}{3}}\right)}{\left(4+675 \alpha ^3 k_J^2+\left|4-675 \alpha ^3 k_J^6\right|\right){}^{\frac{1}{3}}}~.
\end{equation}

Let us consider the case of $f_1(R)=R+aR^2$ and $f_2(R)=1+bR$, for which $a$ and $b$ are related to each other according to the relation $2\alpha\gamma=\beta$. For this choice of functions, we have $\alpha=2a$, $\beta=b/\kappa$ and $\gamma=1/\kappa$, hence $4a=b$. Thus:
\begin{equation}
    b=\frac{10^{2/3}-60 \sqrt[3]{5} \epsilon \kappa  k_J^2}{225 \sqrt[3]{2} \epsilon \kappa  k_J^4}~,
\end{equation}
which for the Bok globules reported in Table \ref{tab:globules}, we get an estimate of $b=\mathcal{O}(-10^{29})~m^2$, which is a too strong coupling constant, hence this specific combination of the functions $f_1$ and $f_2$ seems to be not viable.

}

\section{Conclusions}\label{sec:conclusion}

In this paper, we have studied the phenomenon of Jeans gravitational instability in the context of non-minimal matter-curvature coupling alternative gravity model, in the quantum regime. In the weak-field limit, the model reduces to a modified Schrödinger-Poisson (or Schrödinger-Newton) model, that we have treated kinetically by relying on the use of Wigner functions, allowing to represent quantum states in the classical phase space. Our results generalize those of \cite{Claudio} and enable to study the interplay between non-minimal matter-curvature effects, finite temperature effects, and quantum effects. We have discussed special cases of the model (general relativity, $f(R)$ theories, pure non-minimal
coupling, and $2\alpha \gamma = \beta$) and have used the data of Ref. \cite{Kandori} for Bok globules to constraint the parameters of the model.

Bok globules constitute, in fact, an excellent laboratory to test modified gravity models because their mass is of the same order as their Jeans mass; hence a small deviation from the Jeans criterion leads to a different prediction for their stability. Our approach remains nevertheless valid for compact objects where quantum effects are dominant, opening up novel avenues to test this alternative gravity model. Besides, given the universality of the Schrödinger-Poisson model, which applies equally well to other media, such as plasmas and cold atomic clouds, our results may guide future laboratory experiments to emulate alternative theories of gravity in condensed-matter gravity analogs (see for instance \cite{GR3,NG1}). 

\section*{Acknowledgements}
C.G. is supported by Fundo Regional para a Ciência e Tecnologia and Azores Government Grant No. M3.2DOCPROF/F/008/2020.


\begin{thebibliography}{99}
\bibitem{Will} C. M. Will, {The Confrontation between General Relativity and Experiment}, \href{ 	
https://doi.org/10.12942/lrr-2014-4}{Living Rev. Rel. \textbf{17}, 4 (2014).}
\bibitem{bertolamireview} O. Bertolami, and J. Páramos, {The experimental status of Special and General Relativity} {\it in} Springer Spacetime Handbook (2014), [arXiv:1212.2177 [gr-qc]].
\bibitem{Barack} L. Barack \textit{et al.}, {Black holes, gravitational waves and fundamental physics: a roadmap}, \href{https://doi.org/10.1088/1361-6382/ab0587}{Class. Quantum Grav. \textbf{36}, 143001 (2019).}
\bibitem{mg0} S. M. Carroll, A. De Felice, V. Duvvuri, D. A. Easson, M. Trodden, and M. S. Turner, {Cosmology of generalized modified gravity models}, \href{https://doi.org/10.1103/PhysRevD.71.063513}{Phys. Rev. D \textbf{71}, 063513 (2005).}
\bibitem{mg1} E. J. Copeland, M. Sami, and S. Tsujikawa, {Dynamics of Dark Energy}, \href{https://doi.org/10.1142/S021827180600942X}{Int. J. Mod. Phys. D \textbf{15}, 1753 (2006).}
\bibitem{mg2} S. Nojiri and S. D. Odintsov, {Introduction to Modified Gravity and Gravitational Alternative for Dark Energy}, \href{ 	
https://doi.org/10.1142/S0219887807001928}{Int. J. Geom. Meth. Mod. Phys. \textbf{4}, 115 (2007).}
\bibitem{mg3} S. Nojiri, S.D. Odintsov, V.K. Oikonomou, {Modified Gravity Theories on a Nutshell: Inflation, Bounce and Late-time Evolution}, \href{ 	
https://doi.org/10.1016/j.physrep.2017.06.001}{Phys. Rep. \textbf{692} (2017).} 
\bibitem{mg4} N. D. Birrel and P. C. W. Davies, \textit{Quantum Fields in Curved Space } (Cambridge University Press, Cambridge, England, 1982).
\bibitem{mg5} L. Parker and D. J. Toms, \textit{Quantum Field Theory in Curved Spacetime: Quantized Fields and Gravity} (Cambridge University Press, Cambridge, England, 2009). 
\bibitem{mg6} J. M. M. Senovilla and D. Garfinkle, {The 1965 Penrose singularity theorem}, \href{https://doi.org/10.1088/0264-9381/32/12/124008
}{Class. Quant. Grav. \textbf{32}, 124008 (2015).}
\bibitem{smg1} A. De Felice and S. Tsujikawa, {f(R) theories}, \href{ 	
https://doi.org/10.12942/lrr-2010-3}{Living Rev. Rel. \textbf{13}, 3 (2010).}
\bibitem{smg2} C. H. Brans and R. H. Dicke, {Mach's Principle and a Relativistic Theory of Gravitation}, \href{https://doi.org/10.1103/PhysRev.124.925}{Phys. Rev. \textbf{124}, 925 (1961).}
\bibitem{smg3}  P. G. Bergmann, {Comments on the scalar-tensor theory}, \href{https://doi.org/10.1007/BF00668828}{Int. J. Theor. Phys. \textbf{1}, 25 (1968).}
\bibitem{smg4} J. Beltran Jimenez, L. Heisenberg and T. S. Koivisto, {The Geometrical Trinity of Gravity}, arXiv:1903.06830 [hep-th].
\bibitem{smg5} M. P. Dabrowski and K. Marosek, {Regularizing cosmological singularities by varying physical constants}, \href{ 	
https://doi.org/10.1088/1475-7516/2013/02/012}{JCAP \textbf{1302}, 012 (2013).} 
\bibitem{smg6} K. Leszczynska, A. Balcerzak and M. P. Dabrowski, {Varying constants quantum cosmology}, \href{ 	
https://doi.org/10.1088/1475-7516/2015/02/012}{JCAP \textbf{1502}, 012 (2015).}

\bibitem{f1} T.P. Sotiriou, V. Faraoni, {f(R) theories of gravity}, \href{https://doi.org/10.1103/RevModPhys.82.451}{Rev. Mod. Phys. \textbf{82}, 451 (2010).}
\bibitem{f2} S. Capozziello and M. De Laurentis, {Extended Theories of Gravity}, \href{ 	
https://doi.org/10.1016/j.physrep.2011.09.003
}{Phys. Rep. \textbf{509}, 167 (2011).}

\bibitem{starobinskyinflation}
A.A. Starobinsnky, {A new type of isotropic cosmological models without singularity}, \href{https://doi.org/10.1016/0370-2693(80)90670-X}{Phys. Lett. B {\bf 91}, 99 (1980).}

\bibitem{f3} P.A.R. Ade \textit{et al.} [Planck Collaboration], {Planck 2015 results. XX. Constraints on inflation}, \href{ https://doi.org/10.1051/0004-6361/201525898}{Astron. Astrophys. \textbf{594}, A20 (2016).}
\bibitem{f4} A.A. Starobinsky, {Disappearing cosmological constant in f(R) gravity}, \href{https://doi.org/10.1134/S0021364007150027}{JETP Lett. \textbf{86}, 157 (2007).}



\bibitem{NMC} O. Bertolami, C.G. Böhmer, T. Harko, F.S.N. Lobo, {Extra force in f(R) modified theories of gravity}, \href{https://doi.org/10.1103/PhysRevD.75.104016}{Phys. Rev. D \textbf{75}, 104016 (2007).}
\bibitem{NMC1} O. Bertolami, J. Páramos, {Mimicking dark matter through a non-minimal gravitational coupling with matter}, \href{ 	
https://doi.org/10.1088/1475-7516/2010/03/009}{JCAP \textbf{03}, 009 (2010).}
\bibitem{NMC2} O. Bertolami, P. Frazão, J. Páramos, {Mimicking dark matter in galaxy clusters through a nonminimal gravitational coupling with matter}, \href{https://doi.org/10.1103/PhysRevD.86.044034}{Phys. Rev. D \textbf{86}, 044034 (2012).}
\bibitem{NMC3} O. Bertolami, P. Frazão, J. Páramos, {Accelerated expansion from a nonminimal gravitational coupling to matter}, \href{https://doi.org/10.1103/PhysRevD.81.104046}{Phys. Rev. D \textbf{81}, 104046 (2010).}
\bibitem{NMC4} C. Gomes, J.G. Rosa, O. Bertolami, {Inflation in non-minimal matter-curvature coupling theories}, \href{ 	
https://doi.org/10.1088/1475-7516/2017/06/021}{JCAP \textbf{06}, 021 (2017).}
\bibitem{NMC5} O. Bertolami, C. Gomes, F.S.N. Lobo, {Gravitational waves in theories with a non-minimal curvature-matter coupling}, \href{ 	
https://doi.org/10.1140/epjc/s10052-018-5781-5}{Eur. Phys. J.
C \textbf{87}(4), 303 (2018).}
\bibitem{NMC6} O. Bertolami, C. Gomes, {The Layzer-Irvine equation in theories with non-minimal coupling between matter and curvature}, \href{ 	
https://doi.org/10.1088/1475-7516/2014/09/010}{JCAP \textbf{09}, 010 (2014).}
\bibitem{NMC7} O. Bertolami, A. Martins, {Dynamics of perfect fluids in nonminimally coupled gravity}, \href{https://doi.org/10.1103/PhysRevD.85.024012}{Phys. Rev. D \textbf{85}, 024012 (2011)}.
\bibitem{NMCB} O. Bertolami and C. Gomes, {Nonminimally coupled Boltzmann equation: Foundations}
\href{https://doi.org/10.1103/PhysRevD.102.084051}{Phys. Rev. D \textbf{102}, 084051 (2020).}
\bibitem{Claudio} C. Gomes, {Jeans instability in non-minimal matter-curvature coupling
gravity}, \href{https://doi.org/10.1140/epjc/s10052-020-8189-y}{Eur. Phys. J. C \textbf{80}, 633 (2020).}


\bibitem{Ferreira1} T. D. Ferreira, N. A. Silva, O. Bertolami, C. Gomes, and A. Guerreiro, Simulating N-body systems for alternative theories of gravity using solvers from nonlocal optics, \href{https://doi.org/10.1117/12.2527295}{Proc. SPIE 11207, Fourth International Conference on Applications of Optics and Photonics, 1120710 (3 October 2019).} 



\bibitem{Ferreira2} T. D. Ferreira, N. A. Silva, O. Bertolami, C. Gomes, A. Guerreiro, Using numerical methods from nonlocal optics to simulate the dynamics of N-body systems in alternative theories of gravity, \href{https://doi.org/10.1103/PhysRevE.101.023301}{Phys. Rev. E \textbf{101}, 023301 (2020).} 

\bibitem{Ferreira3} T. D. Ferreira, J. Novo, N. A. Silva, A. Guerreiro, O. Bertolami, Pressureless stationary solutions in a Newton-Yukawa gravity model, \href{https://doi.org/10.1103/PhysRevD.103.124019}{Phys. Rev. D \textbf{103}, 124019 (2021).} 

\bibitem{SN1} L. Diósi, {Gravitation and quantum-mechanical localization of macro-objects}, \href{https://doi.org/10.1016/0375-9601(84)90397-9}{Phys. Lett. A \textbf{105}, 199 (1984).}  
\bibitem{SN2} R. Penrose, {On gravity’s role in quantum state reduction}, \href{https://doi.org/10.1007/BF02105068}{Gen. Relativ. Gravit. \textbf{28}, 581 (1996).} 

\bibitem{BS1} D. J. Kaup, {Klein-Gordon Geon}, \href{https://doi.org/10.1103/PhysRev.172.1331}{Phys. Rev. \textbf{172}, 1331 (1968).} 




\bibitem{BS2} R. Ruffini and S. Bonazzola, {Systems of Self-Gravitating Particles in General Relativity and the Concept of an Equation of State}, \href{https://doi.org/10.1103/PhysRev.187.1767}{Phys. Rev. \textbf{187}, 1767
(1969).} 

\bibitem{BS3} F. E. Schunck and E. W. Mielke, {General relativistic boson stars}, \href{https://doi.org/10.1088/0264-9381/20/20/201}{Classical Quantum Gravity \textbf{20}, R301 (2003).} 

\bibitem{LIGO} N. Sennett, T. Hinderer, J. Steinhoff, A. Buonanno, and
S. Ossokine, {Distinguishing boson stars from black holes and neutron stars from tidal interactions in inspiraling binary systems}, \href{https://doi.org/10.1103/PhysRevD.96.024002}{Phys. Rev. D \textbf{96}, 024002 (2017).} 

\bibitem{SFDM1} M. Membrado, A. F. Pacheco, and J. Sañudo, {Hartree solutions for the self-Yukawian boson sphere}, \href{https://doi.org/10.1103/PhysRevA.39.4207}{Phys. Rev. A \textbf{39}, 4207 (1989).} 



\bibitem{SFDM2} W. Hu, R. Barkana, and A. Gruzinov, {Fuzzy Cold Dark Matter: The Wave Properties of Ultralight Particles}, \href{https://doi.org/10.1103/PhysRevLett.85.1158}{Phys. Rev.
Lett. \textbf{85}, 1158 (2000).} 

\bibitem{SFDM3} P.-H. Chavanis, {Mass-radius relation of Newtonian self-gravitating Bose-Einstein condensates with short-range interactions. I. Analytical results}, \href{https://doi.org/10.1103/PhysRevD.84.043531}{Phys. Rev. D \textbf{84}, 043531 (2011).} 

\bibitem{SFDM4} L. Hui, J. P. Ostriker, S. Tremaine, and E. Witten, {Ultralight scalars as cosmological dark matter}, \href{https://doi.org/10.1103/PhysRevD.95.043541}{Phys. Rev. D \textbf{95}, 043541 (2017).} 


\bibitem{pl0} E. M. Lifshitz, and L. P. Pitaevskii, \textit{Physical Kinetics} (Pergamon Press: Oxford, UK, 1981).

\bibitem{pl1} F. Haas, G. Manfredi, and M. Feix, {Multistream model for quantum plasmas}, \href{https://doi.org/10.1103/PhysRevE.62.2763}{{{Phys. Rev.} E \textbf{62}, 2763 (2020).}} 
 
\bibitem{pl2} G. Manfredi and F. Haas, {Self-consistent fluid model for a quantum electron gas}, \href{https://doi.org/10.1103/PhysRevB.64.075316}{{{Phys. Rev.} B \textbf{64}, 075316 (2001).}} 

\bibitem{MOT1} T. Walker, D. Sesko, and C. Wieman, {Collective behavior of optically trapped neutral atoms}, \href{https://doi.org/10.1103/PhysRevLett.64.408}{Phys. Rev. Lett. \textbf{64}, 408 (1990).} 

\bibitem{MOT2} L. Pruvost, I. Serre, H. T. Duong, and J. Jortner, {Expansion and cooling of a bright rubidium three-dimensional optical molasses}, \href{https://doi.org/10.1103/PhysRevA.61.053408}{Phys. Rev. A \textbf{61}, 53408 (2000).}


\bibitem{MOT3} J. T. Mendonça, R. Kaiser, H. Terças, and J. Loureiro, {Collective oscillations in ultracold atomic gas}, \href{https://journals.aps.org/pra/abstract/10.1103/PhysRevA.78.013408}{Phys. Rev. A \textbf{78}, 013408 (2008).}
\bibitem{ga0} R. Bekenstein, R. Schley, M. Mutzafi, \textit{et al.}, {Optical simulations of gravitational effects in the Newton–Schrödinger system}, \href{https://doi.org/10.1038/nphys3451}{Nature Phys. \textbf{11}, 872 (2015).}  
\bibitem{ga1} J. T. Mendonça, {Wave-kinetic approach to the Schrödinger–Newton equation}, \href{https://doi.org/10.1088/1367-2630/ab0045}{New J. Phys. \textbf{21}, 023004 (2019).}
\bibitem{ga2} K. Ourabah, {Fingerprints of nonequilibrium stationary distributions in dispersion relations}, \href{https://www.nature.com/articles/s41598-021-91455-1}{Sci. Rep. \textbf{11}, 12103 (2021).}






\bibitem{Wigner} E. P. Wigner, {On the Quantum Correction For Thermodynamic Equilibrium}, \href{https://doi.org/10.1103/PhysRev.40.749}{Phys. Rev. \textbf{40}, 749 (1932).}
\bibitem{Moyal} J. E. Moyal, Quantum mechanics as a statistical theory, in \textit{Mathematical Proceedings of the Cambridge Philosophical
Society} (Cambridge University Press, Cambridge, 1949), Vol.
45, pp. 99–124.
\bibitem{Tito} J. T. Mendonça and H. Terças, \textit{Physics of Ultra-Cold
Matter}, Springer Series on Atomic, Optical and Plasma
Physics \textbf{70} (2013).
\bibitem{Our1} K. Ourabah, {Quasiequilibrium self-gravitating systems}, \href{https://doi.org/10.1103/PhysRevD.102.043017}{Phys. Rev. D \textbf{102}, 043017 (2020).}

\bibitem{Our2} K. Ourabah,{
Linear dark matter density perturbations: A Wigner approach}, \href{https://iopscience.iop.org/article/10.1209/0295-5075/132/19002/meta}{EPL \textbf{132}, 19002 (2020).}

\bibitem{Her} A. Hern\'{a}ndez-Almada, M. A. Rodr\'{i}guez-Meza, and T. Matos, {Jeans’ instability analysis of scalar field halos}, \href{https://doi.org/10.1063/1.3647547}{AIP Conference Series \textbf{1396}, 196 (2011).}

\bibitem{Madelung} E. Madelung, {Quantentheorie in hydrodynamischer Form}, \href{https://link.springer.com/article/10.1007/BF01400372}{Z. Phys. \textbf{40}, 322 (1927).}

\bibitem{Bohm} D. Bohm,  {A Suggested Interpretation of the Quantum Theory in Terms of "Hidden" Variables. I}, \href{https://doi.org/10.1103/PhysRev.85.166}{Phys. Rev. \textbf{85}, 166 (1952).} 



\bibitem{Kandori} R. Kandori, et al., {Near Infrared Imaging Survey of Bok Globules: Density Structure}, \href{https://doi.org/10.1086/444619}{Astron.J. {\bf 130}, 2166 (2005).}

\bibitem{Vainio} J. Vainio and I. Vilja, {Jeans analysis of Bok globules in f(R) gravity}, \href{https://doi.org/10.1007/s10714-016-2120-8}{Gen. Rel. Grav. {\bf 48}, 129 (2016).}

\bibitem{nuclearconstraints} S. B. Fisher and E. D. Carlson, {Nuclear Limits on Non-Minimally Coupled Gravity}, \href{https://doi.org/10.1103/PhysRevD.105.024020}{Phys. Rev. D {\bf 105}, 024020  (2022).}


\bibitem{GR3} R. Bekenstein, R. Schley, M. Mutzafi, \textit{et al.}, {Optical simulations of gravitational effects in the Newton–Schrödinger system}, \href{https://doi.org/10.1038/nphys3451}{Nature Phys. \textbf{11}, 872 (2015).}  
\bibitem{NG1}
M. Chalony, J. Barré, B. Marcos, A. Olivetti, and D. Wilkowski, {Long-range one-dimensional gravitational-like interaction in a neutral atomic cold gas}, \href{https://doi.org/10.1103/PhysRevA.87.013401}{Phys. Rev. A \textbf{87}, 013401 (2013).}

\end{thebibliography}
\end{document}